\documentclass{elsart}
\usepackage{graphicx,natbib}
\bibpunct{(}{)}{;}{a}{}{,}
\journal{New Astronomy}
\begin{document}

\begin{frontmatter}

\title{Visible--IR Colors and Lightcurve Analysis of Two Bright TNOs: 
$1999$ TC$_{36}$ and $1998$ SN$_{165}$ \thanksref{titlefn}}
\thanks[titlefn]{Based on observations carried out at the New 
Technology Telescope from the European Southern Observatory (NTT; La 
Silla, Chile) and the Italian Telescopio Nazionale Galileo operated by
 the Centro Galileo Galilei of the CNAA at the Observatorio del Roque 
de los Muchachos (TNG; La Palma, Spain).}

\author{N. Peixinho\thanksref{fn}}
\thanks[fn]{E-mail: Nuno.Peixinho@obspm.fr}
\address{Observatoire de Paris, LESIA, 
F-92195 Meudon Cedex, France \\ and \\
CAAUL, Tapada da Ajuda, PT-1349-018 Lisboa, Portugal}
\author{A. Doressoundiram}
\address{Observatoire de Paris, LESIA, 
F-92195 Meudon Cedex, France}
\author{J. Romon-Martin}
\address{Observatoire de Paris, LESIA, 
F-92195 Meudon Cedex, France}

%
%

\begin{abstract}
We report on observations of two bright Trans-Neptunian Objects (TNOs) 
-- $1999$ TC$_{36}$ and $1998$ SN$_{165}$ -- during two
observational campaigns, as part of the Meudon Multicolor Survey of 
Outer Solar System Objects.
$V-J$ color was measured for $1999$ TC$_{36}$ ($V-J=2.34 \pm 0.18$),
which combined with previous measured colors in the visible, 
indicate a red reflectivity spectrum at all wavelengths.  
Photometric V-band lightcurves were taken for both objects over a time
span of around 8 hours.
We have determined a possible rotational period of $P=10.1 \pm 0.8~h$ 
for $1998$ SN$_{165}$, making it the seventh TNO with an estimated 
period. From its lightcurve variation 
of $\Delta m=0.151^{+0.022}_{-0.030}$, we have inferred an 
asymmetry ratio of $a/b \geq 1.148^{+0.024}_{-0.031}$.
For $1999$ TC$_{36}$, we did not detect any rotational period or 
periodic signal variation within the uncertainties, 
but the analysis of its lightcurve hints to a slight systematic 
magnitude decrease.
\end{abstract}


\end{frontmatter}

\section{Introduction}
Beyond the orbit of Neptune there exists a population of bodies remnant
from the formation of the solar system, planetesimals whose
number density inhibited further accretion and whose orbits
are,  in general,  stable on solar system time scales.  These
are the so-called Trans-Neptunian Objects (TNOs), also known as
 Edgeworth-Kuiper Objects (EKOs) or Kuiper-Belt Objects (KBOs).
Postulated by \cite{Edg43, Edg49} and \cite{Kui51}, their existence
has only recently been confirmed observationally by \cite{JewLuu93}.

The Edgeworth-Kuiper Belt (EKB) is also most probably the source of 
the short-period comets \citep{Dun88}
and a transient population of objects between the orbits of Jupiter and
Neptune: the Centaurs.

Although there are no strict definitions,
the TNOs are generally classified in three groups: resonant objects, 
classical objects and scattered objects.
The resonant are objects trapped in orbital resonances with Neptune, 
mainly in the 2:3 like Pluto, and are therefore also called Plutinos. 
The classical objects have semi-major axes mostly confined between 40 
and 48 AU. Scattered objects have highly eccentric 
and inclined orbits. Presently, about 590 TNOs and 33 Centaurs are 
known.

The physical properties of the TNOs and 
Centaurs are still poorly understood. Spectroscopic studies of these 
objects are only possible with 8-10 meter class telescopes 
and limited to the brightest ones. Therefore, 
broadband photometry is still the most feasible method allowing a 
compositional survey relevant for 
statistical work. At present, several adequate samplings of the 
visible colors of these objects are published, allowing already some 
statistical analysis \citep[e.g.:][ and references therein]{TegRom00,
Dor01,HaiDel02}. However, this data is not yet 
enough to support or refute some claimed relations 
between colors, sizes and orbital parameters. 
On the other hand, \cite{Bar01} showed 
the importance of the $V-J$ color in any taxonomical work 
characterizing the TNOs, but with $\sim 20$ objects measured so far 
\citep[][ and references therein]{Boe01,Dav00} IR data is still scarce.

Precise and unequivocal determination of TNOs' rotational periods is 
difficult to accomplish. \cite{RomTeg99} detected for the first time
lightcurve variability among this class of objects, estimating 
rotational periods for $1995$ QY$_9$, $1994$ VK$_8$ and $1994$ TB. 
Presently, three more TNOs have published periods with good accuracy: 
Varuna \citep{Far01,JewShe02}, $1996$ TO$_{66}$ \citep{Hai00} and 
$1998$ SM$_{165}$ \citep{Roma01}. This latter object has recently been 
found to be a binary \citep{BroTru02}. 
Centaurs, which are brighter and easier to study,  are better sampled, 
with rotational periods 
published for Chiron \citep{Bus89}, Asbolus \citep{BroLuu97}, Pholus 
\citep{BuiBus92},
$1999$ UG$_5$ \citep{Gut01}, $2000$ QC$_{243}$ and $2001$ PT$_{13}$ 
\citep{Ort02}.

Knowledge of rotation rates of TNOs and Centaurs may provide insight
into their collisional evolution. 
Indeed, the present wide color diversity among TNOs may have 
originated from collisional resurfacing processes.
Moreover, albedo measurements, spectroscopic studies, and accurate 
multicolor photometry on these objects depend 
critically on the knowledge of their rotational properties (or 
magnitude variations). This is necessary due to the 
non-simultaneity of the observations in the several bands and the 
long exposure time needed, particularly in the infrared.
Large samples are usually necessary for precise determination 
of the rotational pe\-ri\-ods. Such samples are hard to obtain due to the 
faintness and small magnitude variations expected in the vast majority
 of the TNOs. The identification of 
short term magnitude variations is, consequently, of most importance, 
as it allows us to filter candidates for rotational period 
detections.

We here report on observations of two TNOs. Our results consist of a 
possible determination of a rotational period for $1998$ SN$_{165}$ and 
a measurement of the $V-J$ color for $1999$ TC$_{36}$. Within our 
observational errors we can exclude a rotational period inferior to 8 
hours for $1999$ TC$_{36}$. These two objects are 
of increasing interest, since $1999$ TC$_{36}$ has recently been 
reported to possess a companion object \citep{TruBro02} 
and it has been suggested that $1998$ SN$_{165}$ belongs to a new 
dynamical sub-class of TNOs \citep{Dor02}.
%
%
\section{Observations}
This study is based on two observational campaigns: one dedicated to 
J-band photometry and the other to lightcurve analysis.

J-band photometric data was obtained on the 9th of August 2000, with 
the ARNICA near-infrared camera on the 3.58 m Italian National Telescope 
(TNG; La Palma, Spain) equipped with a HgTeCd array detector with 
$256 \times 256$ pixels (pixel scale=1'', pixel size=40 $\mu m$).

V-band relative photometric data was obtained on the 30th of September 
2000, at the 3.6 m  New Technology Telescope (NTT; ESO, La Silla, 
Chile) with the 
SUSI2 CCD camera on the f/11 Nasmith focus, equipped with a mosaic of 
two EEV CCDs with $2048 \times 4096$ pixels each 
(pixel scale=0.16'', pixel size=15 $\mu m$). This night was 
non-photometric -- seeing varied from 1.3'' to 2.6''.

The sufficiently small angular motion of the targets allowed for 
sidereal tracking rate in all runs (see table \ref{tab:obs_circums}).

   \begin{table}
\setlength{\tabcolsep}{1mm}
{\scriptsize
     \caption{Observational Parameters}
         \label{tab:obs_circums}
\begin{tabular}{ccccccc}
\hline
\hline
Object & Date & R.A. & DEC. & r (AU) & $\Delta$ (AU) & $\alpha$ 
($^{\circ}$) \\
\hline
1998 SN165 & 2000 Sep 30 & 23 38.63 & -00 48.26 & 38.157 & 37.178 & 
0.3 \\
\hline
1999 TC36 & 2000 Aug 9 & 00 10.81 & -08 06.57 & 31.556 & 30.809 & 
1.3 \\
& 2000 Sep 30 & 00 06.43 & -08 39.89 & 31.538 & 30.560 & 0.4 \\
	\end{tabular}
}
	\end{table}
%

\section{Data Reduction}
V-band lightcurve data was reduced with IRAF's CCDRED package. For 
$1999$ TC$_{36}$, we used an aperture radius of 4.5''
centered around the object (and reference stars) with 
the sky level estimated by using the mode value within a concentric 
ring  between 11.2'' and  12.8 '' (farther than 4.3 times the 
worst seeing). Relative magnitude was computed relative to a 
``superstar'' made with the 3 brightest field stars, whose stability  
was verified. To check for the significance 
of eventual magnitude variations on the target, two 
stars of about the same magnitude as the object were also measured. 
For $1998$ SN$_{165}$, an aperture of 3.8'' was used. This object was 
fainter than the previous and more sensitive to sky variations, 
therefore a smaller aperture was used in order to improve the 
S/N ratio. Sky was estimated on a ring between 8.0'' and 9.6'' in most
of the images due to the presence of a close star in some of the frames. 
However, for images with seeings above 2.0'', the sky ring was moved 
away to 11.2'' to ensure that the sky was being 
estimated at distances were no (significant) signal from the object 
was expected. Magnitude was computed relative to a 
``superstar'' made with the 2 brightest field stars 
that were also checked for stability. A star with 
magnitude close to the object's was taken as a reference for 
fluctuations. Relative magnitudes 
obtained are presented in table \ref{tab:tc36_sn165}.

Note that we may expect some flux loss in a few bad seeing images with
 the fixed apertures used. However, since we performed relative 
photometry, this effect will not change the measured relative 
magnitude if the reference stars have the same FWHM as the object. 
We check for all cases and discard any images presenting seeing 
variations above 0.3''.

Near-Infrared J-band data was reduced using both IRAF and MIDAS 
following the standard techniques of flatfielding, using twilight flats, 
estimation of a sky frame, subtraction of the sky from all frames, 
after proper scaling, determination of the offsets between frames based 
on the image headers and final image registration. Details on the 
reduction procedure can be found in \cite{Rom01}.

   \begin{table}
\setlength{\tabcolsep}{2mm}
{\scriptsize
     \caption{Relative Magnitudes of $1999$ TC$_{36}$ (on the left) 
and $1998$ SN$_{165}$ (on the right) at UT date 2000-09-30.}
         \label{tab:tc36_sn165}
\begin{tabular}{cccc|cccc}
\multicolumn{4}{c}{$1999$ TC$_{36}$} & \multicolumn{4}{c}{$1998$ 
SN$_{165}$} \\
\hline
\hline
            UT(h) & Exp(s) & Mag & Error &  UT(h) & Exp(s) & Mag & 
Error \\
            \hline
0.1413 & 120 & 2.296 & 0.030 & 0.2233 & 240 & 3.651 & 0.043 \\
0.3090 & 120 & 2.288 & 0.030 & 0.3794 & 240 & 3.679 & 0.044 \\
0.4961 & 120 & 2.294 & 0.032 & 0.5486 & 240 & 3.734 & 0.040 \\
0.6386 & 180 & 2.319 & 0.022 & 0.7299 & 300 & 3.681 & 0.033 \\
0.9300 & 134 & 2.305 & 0.032 & 1.1677 & 300 & 3.715 & 0.032 \\
1.3865 & 180 & 2.312 & 0.021 & 1.4627 & 300 & 3.739 & 0.037 \\
1.5659 & 180 & 2.312 & 0.022 & 1.6422 & 300 & 3.732 & 0.034 \\
1.7452 & 180 & 2.300 & 0.021 & 1.8236 & 300 & 3.735 & 0.034 \\
2.0697 & 180 & 2.286 & 0.022 & 2.2150 & 300 & 3.700 & 0.032 \\
2.3954 & 180 & 2.268 & 0.021 & 2.4990 & 300 & 3.660 & 0.031 \\
2.6236 & 180 & 2.282 & 0.020 & 2.7492 & 300 & 3.579 & 0.029 \\
2.8832 & 180 & 2.297 & 0.020 & 2.9792 & 300 & 3.596 & 0.031 \\
3.1314 & 180 & 2.295 & 0.020 & 3.2747 & 300 & 3.565 & 0.032 \\
3.6683 & 180 & 2.300 & 0.019 & 3.4986 & 300 & 3.667 & 0.031 \\
3.9371 & 180 & 2.286 & 0.019 & 3.7495 & 300 & 3.590 & 0.028 \\
4.4094 & 180 & 2.318 & 0.019 & 4.0168 & 300 & 3.584 & 0.030 \\
4.6880 & 180 & 2.283 & 0.019 & 4.4973 & 300 & 3.592 & 0.029 \\
4.9345 & 180 & 2.288 & 0.019 & 4.7706 & 300 & 3.658 & 0.031 \\
5.1691 & 180 & 2.269 & 0.020 & 5.0189 & 300 & 3.688 & 0.033 \\
6.0760 & 180 & 2.256 & 0.021 & 5.2530 & 300 & 3.738 & 0.035 \\
6.3066 & 180 & 2.265 & 0.022 & 5.4889 & 300 & 3.676 & 0.034 \\
6.5421 & 180 & 2.274 & 0.022 & 5.6821 & 300 & 3.702 & 0.035 \\
6.7849 & 180 & 2.256 & 0.030 & 6.3979 & 300 & 3.722 & 0.038 \\
6.9755 & 180 & 2.196 & 0.028 & 6.6228 & 300 & 3.767 & 0.044 \\
7.1694 & 180 & 2.269 & 0.026 & 7.0591 & 300 & 3.732 & 0.054 \\
7.4535 & 180 & 2.165 & 0.024 & 7.2462 & 300 & 3.720 & 0.048 \\
7.6719 & 180 & 2.203 & 0.025 & & & & \\
8.0148 & 180 & 2.232 & 0.027 & & & & \\
8.2472 & 180 & 2.219 & 0.028 & & & & \\
8.5034 & 240 & 2.205 & 0.024 & & & & \\
	\end{tabular}
}
	\end{table}
%

\section{Colors and Reflectivities}
The $V-J$ color of $1999$ TC$_{36}$ was calculated using the V magnitude 
reported by 
\cite{Dor01}, which was obtained with only 22 minutes of time difference 
with our J-band data (see table \ref{tab:colors}). 
To compare the spectral behaviors, we quote also the $BVRI$ colors 
for $1999$ TC$_{36}$ and 
$1998$ SN$_{165}$ \citep{Dor01}, and $V-J$ color for $1998$ SN$_{165}$ 
\citep{McB02}.
Relative spectral reflectances at the central 
wavelengths of the broadband filters, normalized to the 
V filter, are plotted in figure 1.

Analysis of figure 1 shows that 
$1999$ TC$_{36}$ possess a red spectrum at all 
studied wavelengths, indicating that its surface may be covered by 
some organic material \citep{Cru89,Tho87}, while 
$1998$ SN$_{165}$ presents an almost flat spectrum at all 
wavelengths. This illustrates the wide color diversity seen among the 
TNOs population, with colors varying from gray ($1998$ SN$_{165}$) to 
red ($1999$ TC$_{36}$).

   \begin{table}
{\scriptsize
     \caption{Colors for $1998$ SN$_{165}$ and $1999$ TC$_{36}$}
         \label{tab:colors}
\begin{tabular}{cccccccc}
\hline
\hline
Object & $V^*$ & $B-V^*$ & $V-R^*$ & $V-I^*$ & $V-J$ & $H_V^*$ & 
$Size(km)^*$ \\
\hline
$1998$ $SN_{165}$ & $21.55 \pm 0.06$ & $0.82 \pm 0.08 $ & $0.33 
\pm 0.08$ & $0.84 \pm 0.08$ & 
$1.27 \pm 0.05^{**}$ & $5.67 \pm 0.06$ & $488$ \\
$1999$ $TC_{36}$ & $20.49 \pm 0.05$ & $0.99 \pm 0.09$ & $0.65 
\pm 0.06$ & $1.37 \pm 0.07$ & 
${\bf 2.34} \pm {\bf 0.18}$ & $5.40 \pm 0.05$ & $552$ \\
Sun & -- & $0.67$ & $0.36$ & $0.69$ & $1.08$ & -- & -- \\
\hline
\multicolumn{8}{l}{$^*$ Values from Doressoundiram et al. (2001)}\\
\multicolumn{8}{l}{$^{**}$ Value from McBride et al. (2002)}

	\end{tabular}
}
 	\end{table}

   \begin{figure}
{\scriptsize
    \resizebox{\hsize}{!}{\includegraphics{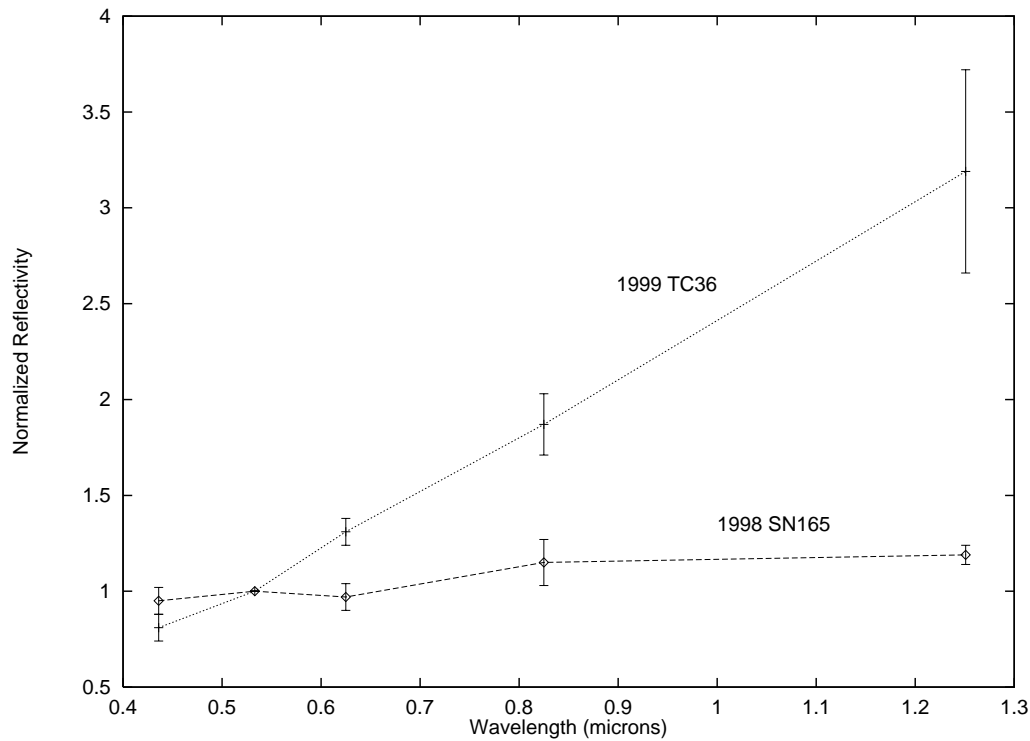}}
      \caption{Relative spectral reflectances of $1998$ SN$_{165}$ 
               and $1999$ TC$_{36}$ normalized to the V filter.
               }
         \label{fig:reflectances}
}
   \end{figure}

\section{Lightcurve Analysis}
\subsection{$1998$ SN$_{165}$}
\begin{figure}
{\scriptsize
    \resizebox{\hsize}{!}{\includegraphics[50,50][316,302]{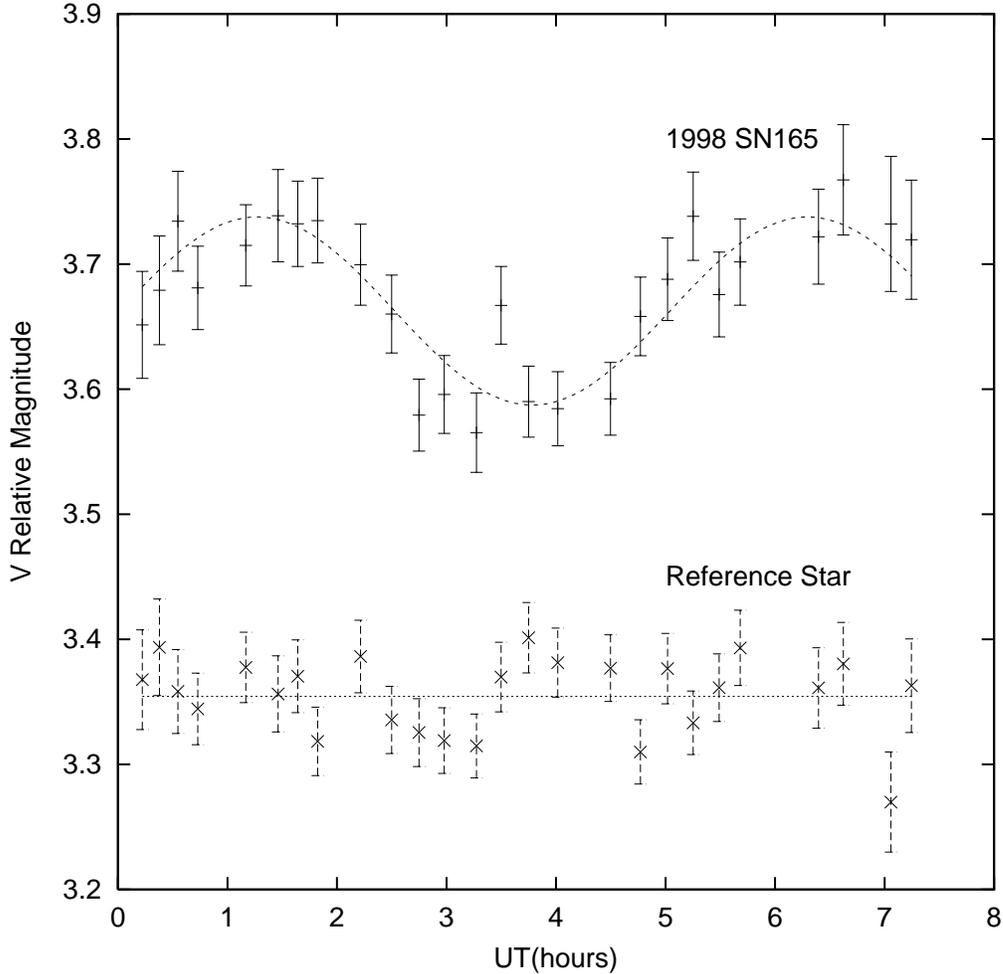}}
      \caption{Lightcurve for $1998$ SN$_{165}$ and reference star 
obtained on 2000-09-30 (UT). X-axis is UT hours and Y-axis is
relative magnitude to the chosen field stars. Magnitudes of the 
reference star were shifted for clarity ($\langle m_v \rangle=3.50$)
               }
         \label{fig:sn165}
}
   \end{figure}

In figure \ref{fig:sn165} we plotted the lightcurve for 
$1998$ SN$_{165}$ resulting from the data in table 
\ref{tab:tc36_sn165}. We have also plotted the data for a reference 
star of about the same magnitude 
as the object, used to check for the significance of any eventual 
variability. It seems patent that 
the object has a periodic variation of magnitude.

A first test for non-random variation was performed with the 
Chi-Square Test, in which the {\it null hypothesis} is: {\it there 
is no non-random variation} \citep{Coll99}. The weighted mean 
magnitude is calculated for the object ($\bar{x}_w$) and then the 
Chi-Square :
$$\chi^2=\sum_{i=1}^n \frac{(x_i - \bar{x}_w)^2}{\sigma_i^2}$$

\noindent
were $x_i$ are the relative magnitude values, $\sigma_i$ the corresponding
 errors and $n$ the number of data-points.
 Assuming that the errors are gaussian, and that $n$ is a large number, 
the mean limiting distribution is $\langle \chi \rangle^2=n-1$ and 
the variance is $\sigma^2=2(n-1)$.
For $1998$ SN$_{165}$ and using a 26 data-point sample, we 
obtained $\chi^2=79.96$,  which is 
$7.8\sigma$ ($\sigma=7.07$) above the mean $\langle \chi \rangle^2=25$.
Since the result is larger than 
$3\sigma$, we have a significant rejection of the null hypothesis. 

\begin{figure}
{\scriptsize
    \resizebox{\hsize}{!}{\includegraphics{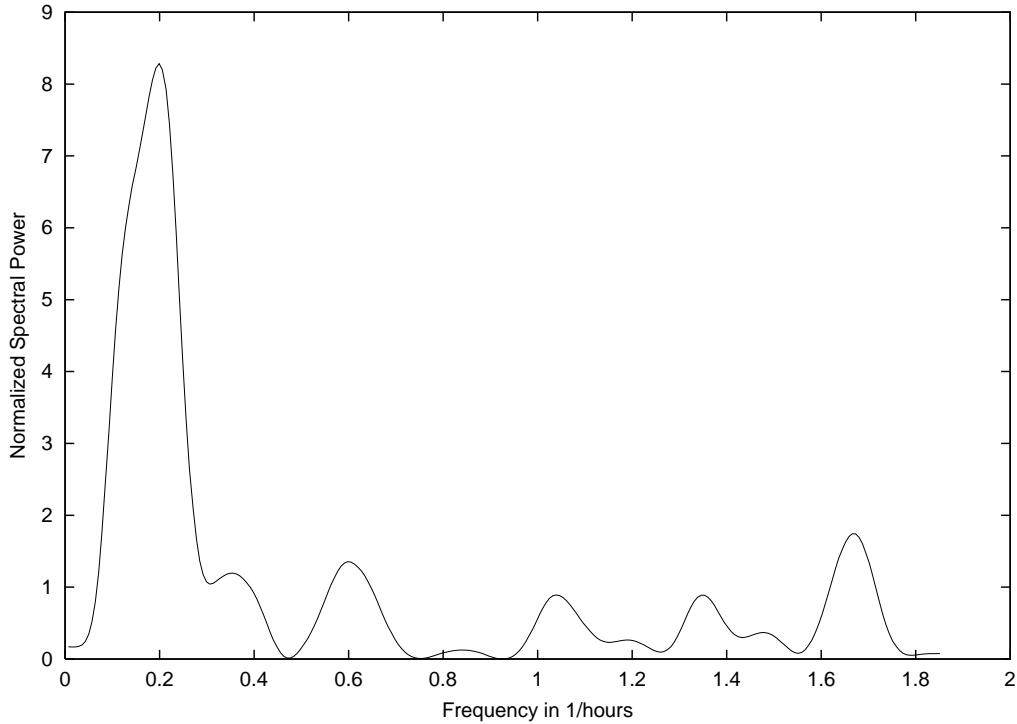}}
      \caption{Lomb-Scargle periodogram for $1998$ SN$_{165}$. 
The frequency is in inverse hours and the power 
relates to the probability that the corresponding 
frequency is due to noise. The higher the power the lower the 
probability. The maximum power of $8.28$ is at $0.199~h^{-1}$.
               }
         \label{fig:lomb_sn165}
}
   \end{figure}

To search for any periodic signal concealed in the data, we applied 
the Lomb-Scargle periodogram \citep{Lom76, Sca82, HorBal86, 
Pre92}. A high peak is detected at the frequency $f=0.199~h^{-1}$
with a power of $8.28$ (fig. \ref{fig:lomb_sn165}).

To check for the level of significance in the estimated period, we 
carried out a Monte-Carlo 
simulation. Holding fixed the number of data-points, their time 
locations $t_i$ and the corresponding errors $\sigma_i$, 
10\,000 random data-sets were constructed.  Each random 
$x_i \equiv x(t_i)$ was generated following a gaussian distribution 
of standard deviation $\sigma_i$.
Applying the Lomb-Scargle periodogram to each one of these random 
data-sets, 
we may estimate the probability of our detected frequency being due 
to noise --- the {\it false period probability} ($P_f$). 
With only 4 events with higher powers than $8.28$ we have a 
$P_f=0.04\%$, that is, a {\it significance level} of $99.96\%$.

By fitting a sine-wave of $f=0.199~h^{-1}$ to the data with the 
Levenberg-Marquardt algorithm, we determine the magnitude variation, 
{\it i.e.}, peak to peak amplitude, of $\Delta m=0.151$. After 
subtracting this signal sine-wave to the data, we obtained a noise 
sample with standard deviation $\sigma_N=0.031$, which is of the order 
of our average error bars. This revealed a good signal subtraction.

To determine the error in the frequency, we also performed a 
Monte-Carlo simulation. Analogously to the previous simulation, we 
generated 1\,000 data-sets, now adding the gaussian noise to the 
signal obtained from the fitted sine wave and applying the 
Lomb-Periodogram to each set. The resulting distribution had a mean 
value of $\langle f\rangle=0.198$ and a standard deviation of 
$\sigma_f=0.016$. Therefore, our detected frequency is: 
$$f=0.199 \pm 0.016~h^{-1}$$

The error in the amplitude is estimated by taking the two extreme 
values (amplitude plus or minus its error from the fit), resulting 
from fitting the sine-wave with $f=0.199 +\sigma_f$ and 
$f=0.199 -\sigma_f$. We get a lightcurve amplitude of:
$$\Delta m=0.151^{+0.022}_{-0.030}$$

However, we should be cautious in considering the significance 
of the detected frequency, since the lightcurve 
is not sufficiently sampled to allow detection of eventually more 
complex behaviors than a single sine-wave variation. Nevertheless, 
we may still make considerations on its rotational period and 
possible asymmetry. Assuming an uniform surface albedo for the object,
 the lightcurve is the result of the variation of the projected 
cross-section area of an elongated shape in rotation. Therefore, a 
full rotation produces a lightcurve with two minima and two maxima. 
With this assumption, the period found with our observations 
is only due to half rotation. Hence, the estimated rotation period is:
$$P=10.1 \pm 0.8~h$$

Supposing a tri-axial ellipsoidal body (with $a>b>c$) rotating along 
the shortest axis ($c$) perpendicular to the line of sight, from 
the determined amplitude of the lightcurve $\Delta m$
we estimate a minimum value for the ratio between the semi-axis $a$ 
and $b$, from the equation $\Delta m \geq 2.5 \log a/b$:
$$\frac{a}{b} \geq 1.148^{+0.024}_{-0.031}$$

However, the hypothesis of a rotating non-uniform albedo object 
producing a mono-peak lightcurve cannot be ruled out, in 
which case the rotational period would be half of the computed value 
above.

In order to check for the real variability of the object's magnitude, 
we performed also the Chi-Square test on a reference star. 
A $\chi^2=27.00$ was found, which is only $0.28\sigma$ above the 
$\langle \chi^2 \rangle$, well within the $3\sigma$ bounds of the 
null hypothesis. The results assure the stability of the relative 
photometry and thus confirm that the object's variability is real.
Also, the Lomb-Scargel periodogram applied to the reference 
star did not reveal any significant periodic signal.

\subsection{$1999$ TC$_{36}$}
\begin{figure}
{\scriptsize
    \resizebox{\hsize}{!}{\includegraphics[50,50][316,302]{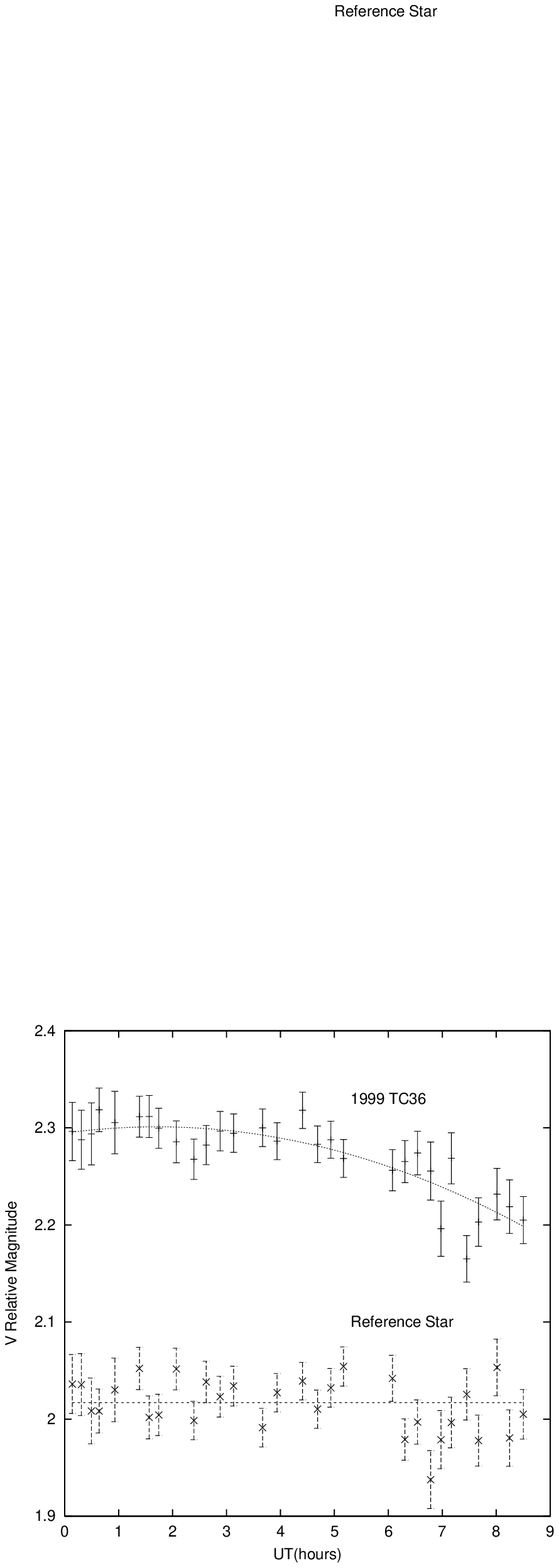}}
      \caption{Lightcurve for $1999$ TC$_{36}$ and reference star 
obtained on the 2000-09-30. X-axis is in UT hours and Y-axis in 
relative magnitude to the chosen field stars. Magnitudes of the 
reference star were shifted for clarity ($\langle m_v \rangle=2.32$).
               }
         \label{fig:tc36}
}
   \end{figure}

We observed $1999$ TC$_{36}$ over 8.362 hours and no apparent periodic
 variation seems present within this time span (see fig. 
\ref{fig:tc36}).

With a sample of 30 data-points, we expect a $\langle \chi^2 
\rangle=29$ and $\sigma=7.62$. The Chi-Square test on the object, 
resulted in a $\chi^2=77.274$, which is $6.34\sigma$ above the 
$\langle \chi^2 \rangle$ . This would imply a 
rejection of the null hypothesis. However, if we reject the 2 lowest 
points that seem slightly wayward, we obtain a $\chi^2=46.78$, now 
only $2.69\sigma$ above $\langle \chi^2 \rangle$ and within 
the null hypothesis interval. Once more, we performed the same test 
on a close magnitude field star (see fig. \ref{fig:tc36}), obtaining 
a $\chi^2=37.24$, only $1.08\sigma$ above $\langle \chi^2 \rangle$. 
This again assures the stability of the 
relative photometry. The Lomb-Scargle periodogram does not find any 
significant period in the data, neither in the object nor in the 
reference star. Comparing the data samples of the object and the 
reference star, 
we see a slight systematic magnitude 
decrease for the object ($\Delta m \approx 0.1$), that might
indicate this object as a good candidate for a longer rotational period. 
However, this
result should be taken with caution since the scatter of data at
the end of the night increases due to the degradation of seeing
conditions. Nonetheless, it is clear that the objects does not possess a 
detectable periodic variation below $8.4~h$ within the observed errors.

\section{Conclusions}
A periodic signal of frequency $f=0.199$ (periodicity $T=5.03~h$) 
with a confidence level above $99.9\%$ is present on our 
V-band lightcurve of $1998$ SN$_{165}$ with a peak-to-peak magnitude 
variation of $\Delta m=0.15$. A possible rotational period of 
$P=10.1 \pm 0.8~h$ and asymmetry ratio of $a/b \geq 1.15$ 
are estimated. This rotational period is to be taken with caution 
and should be confirmed with better sampled observations.
For $1999$ TC$_{36}$, we do not detect any periodic variation 
over our $8.4~h$ time-span  within the uncertainties. If 
$1999$ TC$_{36}$ 
has a detectable rotational period it warrants longer observations.
From the relative reflectance spectra obtained with the BVRIJ colors, 
of $1999$ TC$_{36}$ we see that it possesses a red 
spectrum, probably resulting from a cover 
of organic material on its surface.


%

\section*{Acknowledgments}
The authors most kindly thank text revision from M.E. Filho 
(Kapteyn Inst.) 
and constructive comments from M. Roos-Serote 
(OAL/CAAUL) and M.A. Barucci (Obs. Paris). N.P. is funded by the grant 
(SFRH/BD/1094/2000) and project (ESO/PRO/40158/2000)
from the FCT (Portugal).


\end{document}